\documentstyle[12pt]{article}
\def\abstracts#1#2#3{{
        \centering{\begin{minipage}{4.62in}\baselineskip=13pt
        \small
        \centerline{\bf Abstract}
        \vspace*{0.2cm}
        \parindent=0pt #1\par
        \parindent=18pt #2\par
        \parindent=15pt #3
        \end{minipage} }\par}}
\begin{document}
\vspace*{-2cm}
\hfill \parbox{5cm}{ Mainz preprint\\
                     July 1994 \\
                     October 1994 (rev.) }\\
\vspace*{2cm}
\begin{center}
\LARGE \bf Correlation Function at $\beta_t$ in the Disordered Phase\\
of 2D Potts Models
\end{center}
\vspace*{0.6cm}
\centerline{\large {\em Wolfhard Janke\/} and
                   {\em Stefan Kappler\/}}
\vspace*{0.4cm}
\centerline{\large {\small Institut f\"ur Physik,
                    Johannes Gutenberg-Universit\"at Mainz}}
\centerline{    {\small Staudinger Weg 7, 55099 Mainz, Germany }}
\vspace*{0.5cm}
\abstracts{}{
We use Monte Carlo simulations to measure the spin-spin correlation function 
in the disordered phase of two-dimensional $q$-state Potts models with 
$q=10,15$, and $20$ at the first-order transition point $\beta_t$. 
To extract the correlation length $\xi_d$ from the exponential decay of 
the correlation function over several decades with the desired accuracy
we make extensively use of cluster-update techniques and improved estimators.
Our results for $\xi_d$ are compatible with an analytic formula.
As a byproduct we also measure the energy moments in the disordered phase
and find very good agreement with a recent large $q$ expansion at $\beta_t$.
}{}
\vspace*{1.5cm}
\thispagestyle{empty}
\newpage
\pagenumbering{arabic}
%
                     \section{Introduction}
%
An important quantity to characterize the properties of a statistical system
is the correlation length $\xi$ which can be extracted from the exponential
decay of a correlation function $G(x)$ in the limit of large distances $x$.
Usually various definitions of $G(x)$ are possible and it is a priori unclear
which one is best suited in numerical Monte Carlo simulations. There are only 
very few models for which the correlation length is exactly known and can thus
serve as a testing ground for the employed numerical techniques. The best known
example is the two-dimensional Ising model where the correlation length is
exactly known at all temperatures in both the high- and low-temperature 
phase \cite{ising}. 
Only quite recently also for two-dimensional $q$-state Potts models on simple 
square lattices an analytic formula for the correlation length could be 
derived \cite{xi1,xi2}.
Here, however, the correlation length is only known at one special temperature,
namely at the first-order transition point $\beta_t$ of this model for 
$q \ge 5$. More precisely, by comparing with a large $q$ expansion, it could 
be argued \cite{boja} that the analytic result in Ref.\cite{xi1} refers to the
correlation length 
$\xi_d(\beta_t)$ in the disordered phase. 

Using exact duality arguments and 
the (weak) assumption of complete wetting (which can only be proven in the 
limit of large $q$) this result was then converted into an explicit expression
for the order-disorder interface tension, $\sigma_{od} = 1/2\xi_d$ 
\cite{boja}. This formula turned out to be in good agreement with previous 
(and thus completely unbiased) numerical interface-tension data for 
$q=7$ \cite{ours} and $q=10$ \cite{bn92}, and also subsequent high-precision 
studies obtained compatible values \cite{stlouis}. The purpose of this note 
is to present direct numerical tests of the formula for $\xi_d(\beta_t)$. 
%
           \section{The model and observables}
%
The two-dimensional $q$-state Potts model is defined by the 
partition function \cite{wu}
\begin{equation}
Z = \sum_{\{s_i\}} e^{-\beta E}; \,\,\, E = -\sum_{\langle ij \rangle}
\delta_{s_i s_j}; \,\,\, s_i = 1,\dots,q,
\label{model}
\end{equation}
where $i = (i_x,i_y)$ denote the lattice sites of a square lattice of size
$V = L_x \times L_y$, 
$\langle ij \rangle$ are nearest-neighbor pairs and 
$\delta_{s_i s_j}$ is the Kronecker delta symbol. 
In the infinite volume limit
this model exhibits on simple square lattices for $q\le 4$ ($q \ge 5$) a 
2nd (1st) order phase transition at $\beta_t\!=\!\ln (1+\sqrt{q})$. At 
$\beta_t$ also the internal energy densities $e_o$ and $e_d$ of the ordered 
and disordered phase are known exactly while for the corresponding specific 
heats only the difference $\Delta c = c_d - c_o$ could be derived analytically.

In the disordered phase the spin-spin correlation function can be defined as
\begin{equation}
G(i,j) \equiv \langle \delta_{s_i s_j} - \frac{1}{q} \rangle.
\label{eq:G}
\end{equation}
For numerical purposes it is more convenient to consider the 
$k_y = 0$ projection of $G$,
\begin{equation}
g(i_x,j_x) = \frac{1}{L_y} \sum_{i_y,j_y} G(i,j),
\label{eq:g}
\end{equation}
which should be free of power-like prefactors in the large-distance behaviour.
For periodic boundary conditions translational invariance implies that $g$ 
depends only on $|i_x - j_x|$, and for convenience we shall sometimes simply 
write $g(x)$. 
A useful test of the consistency of our data is provided by the magnetic
susceptibility
\begin{equation}
\chi = \frac{1}{V(q-1)} \langle [\sum_{i} (q\delta_{s_i,1} -1)]^2 \rangle,
\label{eq:chi_def}
\end{equation}
which can be computed from the area under the correlation function,
\begin{equation}
\chi = \frac{q}{V(q-1)} \sum_{i,j} G(i,j) = 
       \frac{q}{q-1} \sum_{i_x=1}^{L_x} g(i_x,0).
\label{eq:chi_g}
\end{equation}
%
                       \section{The simulation}
%
In our Monte Carlo study we investigated the correlation function in the 
disordered phase at $\beta_t$ for $q=10, 15$ and $20$ on lattices 
of size $V = L \times L$  and $V = 2L \times L$ with $L = 150, 60$ and $40$
($\approx 14 \xi_d$). 
To take advantage of translational invariance we used periodic
boundary conditions but chose the lattice sizes large enough to suppress
tunneling events. This guaranteed that, starting from a completely random 
configuration, the system remained a sufficiently long time in the disordered 
phase to perform statistically meaningful measurements. Since in this
situation it is not obvious which update algorithm performs best
we first performed for the $L \times L$ lattices a quite elaborate efficiency 
study of the most
popular update algorithms, the local Metropolis and heat-bath algorithms and
the non-local Wolff single-cluster \cite{wolff} and Swendsen-Wang 
multiple-cluster \cite{sw} algorithms. By measuring
the integrated autocorrelation times $\tau_{{\rm int},e}$ of the energy it 
became immediately clear that the Metropolis algorithm is not a good 
candidate; see Table~1. Also the multiple-cluster algorithm seems to be
inferior in this application. The other two algorithms, on the other hand, 
exhibit comparable $\tau_{{\rm int},e}$, in particular for large $q$ where the
average cluster size is small. Taking into account the details of our 
implementation, this makes it difficult to decide between the two alternatives 
on the basis of $\tau_{{\rm int},e}$ alone. Being mainly interested in the 
long-distance behaviour of correlation functions we therefore also looked at 
the integrated autocorrelation times $\tau_{{\rm int},g(x)}$ of these 
quantities. Our results for $q=20$ are shown in Fig.~1. 
As expected we find that for the local heat-bath algorithm the
autocorrelations grow with distance, while for the non-local single-cluster
algorithm they decrease. On the basis of these tests we finally decided to use
the single-cluster algorithm for all production runs. It should be mentioned
that for any algorithm we used the multiple-cluster decomposition of a given
spin configuration for measurements using the improved estimator
\begin{equation}
G(i,j) = \frac{q-1}{q} \langle \Theta(i,j) \rangle,
\label{eq:G_imp}
\end{equation}
where $\Theta(i,j)=1$, if $i$ and $j$
belong to the same cluster, and $\Theta = 0$ otherwise. By performing the
summations in eq.(\ref{eq:g}) one easily derives an improved estimator 
for $g(i_x,0)$.

In the production runs we updated the spins after many single-cluster iterations
with one multiple-cluster step to facilitate the most efficient use of 
the ``improved estimator'' (\ref{eq:G_imp}). In units of $\tau_{{\rm int},e}$ 
the run time
on the $L \times L$ ($2L \times L$) lattices was about 35\,000 (60\,000) for
$q=10$, 116\,000 (230\,000) for $q=15$, and 72\,000 (35\,000) for $q=20$.
All error bars are estimated by means of the jack-knife technique \cite{jack}.
Finally it is worth mentioning that all our correlation function data are
stored in such a way that they can be reweighted to nearby temperatures in
both directions; in this way we have also computed extrapolations of the
correlation length into the metastable disordered region \cite{inprogress}.
%
                      \section{Results}
%
\subsection{Energy moments}
To convince ourselves that the system was always in the disordered
phase, we monitored the time series of the energy measurements and computed
the first three moments of the energy distribution,
$e_d \equiv \langle E \rangle/V$,
$c_d = \beta_t^2 \mu_d^{(2)} \equiv 
\beta_t^2 ( \langle E^2 \rangle - \langle E \rangle^2)/V$, and
$\mu_d^{(3)} = \langle (E - \langle E \rangle)^3 \rangle /V$. 
While $e_d$ can be compared with exact results, $c_d$ and $\mu_d^{(3)}$ 
can be related by duality to the corresponding moments in the ordered
phase,
\begin{eqnarray} 
c_d &=& c_o + \beta_t^2 (e_d - e_o)/\sqrt{q},\\
\mu_d^{(3)} &=& - \mu_o^{(3)} + 2(1-q)/q^{3/2} + 3 (e_d-e_o)/q
+ 6c_o/\beta_t^2\sqrt{q}, 
\label{eq:dis_ord}
\end{eqnarray}
which have recently been estimated by means of Pad\'e extrapolations of 
large $q$ series expansions \cite{large_q}. Our Monte Carlo estimates for the 
$L \times L$ and $2L \times L$ lattices can be found in
Table~2, together with the Pad\'e extrapolations as given in the reanalysis of
Ref.\cite{private} (using series expansions extended by one term),
which are practically indistinguishable from our own Pad\'e analysis. A 
comparison of the two sets of numbers shows excellent agreement between the 
two methods, even for the third moment and small $q$. 

Estimates of $c_d$ from the finite-size scaling behaviour of the 
specific-heat maxima gave a consistent value of 6.0(2) \cite{fss20} for $q=20$,
but much too small values for $q=10$ \cite{fss10a,fss10b}, while
recent estimates from very long high-temperature series expansions 
\cite{guttmann} are too large by a factor of about 2. Only finite-size 
scaling {\em at\/} the transition point $\beta_t$ seems to 
give sensible results, at least for $q=10$ \cite{fss10b}.
\subsection{Susceptibility}
As a further test of the consistency of our data we compared the magnetic
susceptibility computed according to eq.(\ref{eq:chi_g}) with measurements 
using the improved cluster estimators
\begin{equation}
\chi = \langle |C| \rangle_{SC}
     = \langle |C|^2 \rangle_{SW}/\langle |C| \rangle_{SW},
\label{chi2}
\end{equation}
where $\langle\cdot\rangle_{SC}$ ($\langle\cdot\rangle_{SW}$) refers to the 
average taken from the single (multiple) cluster update. As is shown in 
Table~\ref{tab2}, in all cases we obtained 
excellent agreement between the three estimators. 
\subsection{Correlation function}
Let us now turn to the main subject of this note, the correlation function.
A preliminary report of a first set of $L \times L$ data was recently given
in Ref.\cite{dallas}. Our complete set of measurements now consists of
data for $G$ along the coordinate axes and for the projected correlation 
functions $g(x)$ for $q=10,15$, and $20$ on $L \times L$ and $2L \times L$ 
lattices with $L=150, 60$ and $40$. The average of the $k_y=0$ and $k_x=0$
projections on the $L \times L$ lattices and the $k_y=0$ projection on the
$2L \times L$ lattices, i.e. $g(x)$, are shown in the semi-log plots of 
Fig.~2. The quite pronounced curvature for 
small $x$ indicates that the simplest two-parameter Ansatz for periodic 
boundary conditions,
\begin{equation}
g(x) = a \,{\rm ch}(  \frac{L/2-x}{\xi_d} ),
\label{eq:fit_2}
\end{equation}
which takes into account only the lowest excitation (largest correlation 
length), can only be justified for very large $x$. 
We have therefore considered also the more general Ansatz
\begin{equation}
g(x) = a \,{\rm ch}(   \frac{L/2-x}{\xi_d} ) 
           +   b \,{\rm ch}( c \frac{L/2-x}{\xi_d} ),
\label{eq:fit_4}
\end{equation}
with four parameters $a,b,c$, and $\xi_d$. 

Since non-linear four-parameter fits are notoriously difficult to control, we 
first fixed $\xi_d$ at its theoretical value ($\xi_d=10.559519...$, 
$4.180954...$, and $2.695502...$ for $q=10,15$, and $20$, respectively),
and optimized only the remaining three parameters. The resulting fits
to the $L \times L$ and $2L \times L$ data are shown in Fig.~2 as dotted and 
solid lines, respectively. Over a wide range up to about
$x \approx (5 \dots 6) \xi_d$ the lines are excellent interpolations of the 
data. At very large distances, 
however, we also see a clear tendency of the fits to lie systematically above 
the data. This already indicates that unconstrained fits to the Ansatz 
(\ref{eq:fit_4}) over the same $x$ range with $\xi_d$ as a {\em free} parameter
should somewhat underestimate the analytical value of $\xi_d$. 

In fact, this is what we observed in the unconstrained fits to both the 
$L \times L$ and $2L \times L$ data. 
In order to estimate systematic errors we performed fits to 
both Ans\"atze using varying fit intervals. As a general tendency we noticed 
a trend to higher values for $\xi_d$ when restricting the fit
interval to larger $x$ values, but then also the statistical errors grow
rapidly. For $q=10$ this is illustrated in Fig.~3(a), where $x_{\rm min}$
denotes the smallest $x$ value included in the fits. The last point used was
$x_{\rm max} = L/2$ for both geometries. For the four-parameter
fits we have stopped increasing $x_{\rm min}$ as soon as the error on the
amplitude $b$ became comparable with its central value. For a reasonable
range of $x_{\rm min}$ values satisfying this criterion our results are
collected in Table~4. Here we also give the results of fits of $g(y)$, i.e.,
the $k_x = 0$ projection
along the short direction of the $2L \times L$ lattices. The fits of $g(x)$
with the smallest $x_{\rm min}$ values are shown
in Fig.~2 as long ($L \times L$) and short 
($2L \times L$) dashed lines. For the parameter $c$ we obtain from the 
unconstrained four-parameter fits the $q$ independent estimates of 
$c \approx 1.5 - 2$, with a clear tendency of decreasing $c$ for increasing
$x_{\rm min}$. This observation is consistent with 
the constrained three-parameter fits with $\xi_d$ held
fixed at its analytical value where we find the quite stable estimate of
$c = 1.5 \pm 0.1$, again for all three values of $q$.

Our numerical estimates for $\xi_d$ underestimate the analytical values by 
about $10-20\%$ for both lattice geometries. 
The relative deviation clearly increases with
increasing $q$. For some fit ranges we have repeated the analysis using
so-called correlated fits \cite{michael} which, in general, seemed to be a 
little more stable. We did not observe, however, any significant increase 
of the estimates for $\xi_d$. We also investigated whether the Fourier 
transforms of $g$ or $G$ are less susceptible to systematic corrections and 
thus easier to analyze. Unfortunately, the answer is no. In fact, the fitted 
values of $\xi_d$ turn out to be even smaller than in the corresponding real 
space fits (if comparable fit intervals are used).

Of course, the problem is that at the distances we have
studied so far ($x_{\rm max} = L/2 \approx 7\xi_d$)
even higher excitations cannot be neglected. Due to
convexity properties it is then natural that $\xi_d$ is underestimated 
by using the truncated Ansatz (\ref{eq:fit_4}). This general trend is 
illustrated in another way in Fig.~3(b) where we plot for $q=10$ an 
effective correlation length defined by the local slopes of $g(x)$,
\begin{equation}
\xi_d^{\rm eff} = 1/\ln\left[ g(x)/g(x+1) \right],
\label{eq:xi_eff}
\end{equation}
as a function of the distance $x$ for both the $L \times L$ as well as the
$2L \times L$ data. For large $x$ we expect $\xi_d^{\rm eff} = \xi_d$. We
do oberserve a clear increase of $\xi_d^{\rm eff}$, but it is of course still 
a long way to $\xi_d = 10.56$. In particular with the $L \times L$ data is
is difficult to extrapolate to the correct value since the effects of
the periodic boundary conditions set in much too early. For the $2L \times L$
data, on the other hand, the three-parameter fit (with $\xi_d$ held fixed at 
its theoretical value) indicates how the data should behave for very large 
distances in the long direction of the lattice. At $x=100$, however, 
$g(x)/g(0) \approx 5 \times 10^{-6}$ which is very difficult to measure 
accurately, even with cluster algorithms and improved observables. In fact, 
this number reflects how improbable it is to generate a cluster with diameter 
of about 100 (recall the improved estimator (\ref{eq:G_imp})).
To cope with this problem we are currently investigating a special type of 
simulation with a reweighted Hamiltonian designed to increase these 
probabilities. Using standard simulation techniques it would take an
enormous amount of computing time to follow the decay of correlation functions
over more than 5 or 6 decades with the necessary accuracy.

As a check of our analysis we put $q=2$ in our programs, and thus simulated 
the Ising model in the disordered phase at 
$\beta=0.71 \approx 0.80 \beta_c$. Here the exactly known correlation length is
$\xi_d = 2.728865\dots$ \cite{ising}, a value that is comparable to 
$\xi_d(\beta_t)$ of the $q=20$ Potts model.
Our data points for $g(x)$ on $L \times L$ and $2L \times L$ lattices 
with $L=40$
shown in Fig.~4 look perfectly straight in a semi-log plot. Consequently, the 
much simpler fit of the form (\ref{eq:fit_2}) was 
sufficient. As a result we obtained from the fits (with $x_{\rm min} = 1)$ 
shown in Fig.~4 the estimates of
$\xi_d=2.7232(35)$  ($L \times L$) and $\xi_d=2.7275(24)$ 
($2L \times L$), and for a fit in the short direction of the $2L \times L$ 
lattice $\xi_d=2.7283(20)$. All these estimates are
in very good agreement with the theoretical value, showing that the employed
techniques work at least in principle. 
\section{Discussion}
Previous numerical estimates of the correlation length at $\beta_t$ for
$q=10$ \cite{peczak,fernandez,gupta} resulted in values of about 
$\xi \approx 6$ which are much smaller
than the theoretical prediction of $\xi_d = 10.56$. 
The Fernandez {\em et al.} \cite{fernandez}
value of $6.1(5)$ is obtained from an extrapolation of simulations at
$\beta < \beta_t$ and thus definitely refers to the correlation length
in the disordered phase. From our experience with correlation function fits
and direct tests we believe that their values of $\xi_d$ for $\beta < \beta_t$
are already underestimated. Since the simulation points are relatively
far away from $\beta_t$, the systematic errors are further enhanced by the
extrapolation procedure used in Ref. \cite{fernandez}.
The interpretation of the data of
Peczak and Landau \cite{peczak} and Gupta and Irb\"ack \cite{gupta}
is less clear to us. By repeating the simulations of Ref.\cite{gupta} we are
quite convinced that their technique yields a weighted average of the
ordered and disordered correlation function which is then analyzed to obtain
$\xi$. By using a projection to a momentum $k_1 = 2\pi/L$, the ordered phase
is treated properly, but the weighted average makes the final interpretation
somewhat unclear. Similarly, since the simulations in Ref.\cite{peczak} are
performed at the specific-heat maximum whose finite-size scaling behaviour is 
governed by the transitions between the ordered and disordered phase, it is 
very unlikely that their $\xi$ refers to a pure phase correlation length. 
In view of these problems it is astonishing that all three methods
yield about the same value for $\xi$.
In order to understand this puzzling coincidence 
we are currently investigating also the correlation length in the
ordered phase and first results 
will be available soon in a separate publication \cite{tobe}.

Constrained fits with $\xi_d(\beta_t)$ held fixed at its theoretical value
clearly indicate that our data for the projected correlation function $g(x)$ 
in the disordered phase at $\beta_t$ are compatible with the analytical 
prediction of Refs.\cite{xi1,xi2}. By performing unconstrained fits, however, 
we cannot really 
confirm the theoretical values. Rather we systematically underestimate 
$\xi_d$ by about $10-20\%$ in simulations of $L \times L$ as well as
$2L \times L$ lattices. We attribute this to
higher mass excitations which cannot be neglected at the distances
investigated so far. To include these corrections in the fits, however, would 
require much more precise data. In a comparative study of 2D Ising correlation
functions no such problems were encountered and the exact value of the
correlation length could be reproduced with high precision.
\section*{Acknowledgements}
We would like to thank R. Lacaze and A. Morel for interesting discussions
on the energy moments and for communicating their unpublished results.
WJ gratefully acknowledges a Heisenberg fellowship by the DFG and SK thanks 
the Graduiertenkolleg ``Physik and Chemie supramolekularer Systeme'' for 
financial support. The Monte Carlo simulations were performed on the 
CRAY Y-MP of the H\"ochstleistungsrechenzentrum J\"ulich,
the CRAY Y-MP's of the Norddeutscher Rechnerverbund in Kiel and Berlin,
and on the cluster of fast RISC workstations at Mainz.
%

%
%
\newpage
%
%
\begin{table*}              
{\Large\bf Tables}\\[1cm]
 \begin{center}
  \caption{Integrated autocorrelation times of the energy on $L \times L$
           lattices at $\beta_t$ in the disordered phase for different update 
           algorithms in units of sweeps. The results for Wolff's 
           single-cluster update are rescaled to these units.}
\vspace*{0.5cm}
  \begin{tabular}{|r|r|r|r|}
\hline
\multicolumn{1}{|r|}{Algorithm} &
\multicolumn{1}{c|}{$q=10$}       &
\multicolumn{1}{c|}{$q=15$}       & 
\multicolumn{1}{c|}{$q=20$}       \\
\hline
Metropolis     & $\approx 2000$ & 400(100) & -~~~~ \\
Heatbath       & 125(25)        &  19(5)   & 11(1) \\
Swendsen-Wang  & 175(20)        &  -~~~~   & 67(9) \\
Wolff          & 52(7)          &  23(2)   & 17(2) \\
\hline
\end{tabular}
\end{center}
\end{table*}
%
%
\begin{table}[htb]
\caption[a]{\label{tab:moments}
Comparison of numerical and analytical results for
energy moments at $\beta_t$ in the disordered phase.}
\begin{center}
\begin{tabular}{|l|l|l|l|}
\hline
\multicolumn{1}{|c}{Observable} &
\multicolumn{1}{|c}{$q=10$} &
\multicolumn{1}{|c}{$q=15$} &
\multicolumn{1}{|c|}{$q=20$} \\
\hline
         $e_d\quad$(MC, $\;\:L \times L$) &$ -0.96812(15) $&$ -0.75053(13)  $&$ -0.62648(20)  $\\
         $e_d\quad$(MC, $2L \times L$)&$ -0.968190(81)$&$ -0.750510(65) $&$ -0.626555(97) $\\
         $e_d\quad$(exact)            &$ -0.968203... $&$ -0.750492...  $&$ -0.626529...  $\\
\hline
         $c_d\quad$(MC, $\;\:L \times L$) &$ 18.33(17) $&$ 8.695(47) $&$ 6.144(43) $\\
         $c_d\quad$(MC, $2L \times L$)&$ 18.34(12) $&$ 8.665(29) $&$ 6.140(27) $\\
         $c_d\quad$(large $q$)        &$ 18.5(1)   $&$ 8.66(3)   $&$ 6.133(5)  $\\
\hline
         $\mu_d^{(3)}\;$(MC, $\;\:L \times L$)  &$ -2010(100)$&$ -171.0(5.1)$&$ -54.7(1.9) $\\
         $\mu_d^{(3)}\;$(MC, $2L \times L$) &$ -2031(73) $&$ -176.1(3.8)$&$ -53.9(1.5) $\\
         $\mu_d^{(3)}\;$(large $q$)         &$ -1833(40) $&$ -174(4)    $&$ -54.6(4)   $\\
\hline
\end{tabular}
\end{center}
\end{table}
%
%
\begin{table}[htb]
\caption[a]{\label{tab2} The magnetic susceptibility at $\beta_t$ in the 
                         disordered phase from three different estimators.}
\begin{center}
\begin{tabular}{|c|l|l|l|l|}
\hline
 \multicolumn{1}{|c}{Lattice}       &
 \multicolumn{1}{|c}{Observable}       &
 \multicolumn{1}{|c}{$q=10$} &
 \multicolumn{1}{|c}{$q=15$} &
 \multicolumn{1}{|c|}{$q=20$} \\
\hline
             &$\frac{q}{q-1}\sum^{L}_{i=1}g(i,0)$ &$ 38.02(14)  $&$ 10.228(19)  $&$ 5.874(11)  $\\
$L\times L$&$\langle |C|   \rangle_{SC}$    &$ 38.02(14)  $&$ 10.234(19)  $&$ 5.872(11)  $\\
             &$\langle |C|^2 \rangle_{SW} / \langle |C| \rangle_{SW} $
                                             &$ 38.02(14)  $&$ 10.228(19)  $&$ 5.874(11)  $\\
\hline
              &$\frac{q}{q-1}\sum^{2L}_{i=1}g(i,0)$&$ 38.075(80) $&$ 10.2330(94) $&$ 5.8813(60) $\\
$2L\times L$&$\langle |C|   \rangle_{SC}$   &$ 38.094(80) $&$ 10.2331(91) $&$ 5.8808(59) $\\
              &$\langle |C|^2 \rangle_{SW} / \langle |C| \rangle_{SW} $
                                             &$ 38.075(80) $&$ 10.2330(94) $&$ 5.8813(60) $\\
\hline
\end{tabular}
\end{center}
\end{table}
%
%
%
\begin{table*}              
 \begin{center}
  \caption[a]{Numerical estimates of the correlation length 
              $\xi_d(\beta_t)$ from four-parameter fits to the
              Ansatz (\ref{eq:fit_4}) in the range $x_{\rm min} \dots
              x_{\rm max} = L/2$. For the $2L \times L$ lattices the fits
              along the $x$ and $y$ direction are distinguished by the
              index.}
  \vspace*{0.5cm} 
  \begin{tabular}{|c|cr|cl|cr|}
\hline
\multicolumn{1}{|c|}{ } &
\multicolumn{2}{c|}{$q=10$}       &
\multicolumn{2}{c|}{$q=15$}       &
\multicolumn{2}{c|}{$q=20$}       \\
\multicolumn{1}{|c|}{ } &
\multicolumn{2}{c|}{$L=150$}      &
\multicolumn{2}{c|}{$L=60$}       &
\multicolumn{2}{c|}{$L=40$}       \\
\hline
Lattice & $x_{\rm min}$ & \multicolumn{1}{c|}{$\xi_d$} 
   & $x_{\rm min}$ & \multicolumn{1}{c|}{$\xi_d$}
   & $x_{\rm min}$ & \multicolumn{1}{c|}{$\xi_d$} \\
\hline
                   & 11 & 8.8(3) & 5 & 3.60(10)& 3 & 2.21(6) \\
 $L \times L$      & 16 & 8.9(4) & 7 & 3.67(13)& 4 & 2.21(7) \\
                   & 20 & 9.0(5) & 9 & 3.70(16)& 5 & 2.24(6) \\ 
\hline
                   & 11 & 9.0(4) & 5 & 3.52(5) & 3 & 2.21(3) \\
 $(2L \times L)_x$ & 16 & 9.5(6) & 7 & 3.54(7) & 4 & 2.23(4) \\
                   & 20 & 10.2(9)& 9 & 3.59(10)& 5 & 2.23(5) \\
\hline
                   & 11 & 8.9(4) & 5 & 3.52(8) & 3 & 2.26(4) \\
 $(2L \times L)_y$ & 16 & 9.1(5) & 7 & 3.58(11)& 4 & 2.30(6) \\
                   & 20 & 9.3(7) & 9 & 3.62(16)& 5 & 2.33(7) \\
\hline
exact & \multicolumn{2}{|r|}{$10.559519...$} & 
        \multicolumn{2}{|r|}{$4.180954...$}  &
        \multicolumn{2}{|r|}{$2.695502...$} \\
\hline
\end{tabular}
\end{center}
\end{table*}
%
%
\clearpage
\newpage
{\Large\bf Figure Headings}
  \vspace{1in}
  \begin{description}
    \item[\tt\bf Fig. 1:]
Integrated autocorrelation times of $g(x)$ for $q=20$.
    \item[\tt\bf Fig. 2:]
Semi-log plots of the correlation functions $g(x)$ vs distance $x$ on
$L \times L$ and $2L \times L$ lattices for 
(a) $q=10$, (b) $q=15$, and (c) $q=20$. The solid and dotted lines are 
three-parameter fits to the Ansatz (\ref{eq:fit_4}) with $\xi_d$ held fixed 
at its theoretical value. The short and long dashed lines show unconstrained 
four-parameter fits over the same $x$ range. For clarity some data points are
discarded. 
    \item[\tt\bf Fig. 3:]
(a) Results for $\xi_d$ of the various fits for $q=10$ using all  
data points with $x_{\rm min} \le x \le x_{\rm max}=L/2$ as a function 
of $x_{\rm min}$.\\
(b) The effective correlation length (\ref{eq:xi_eff}) vs distance for $q=10$. 
The dashed lines are constrained three-parameter fits to the data and the
horizontal line shows the theoretically expected result for $\xi_d$.
    \item[\tt\bf Fig. 4]
Semi-log plot of the correlation function $g(x)$ of the
2D Ising model at $\beta = 0.71 \approx 0.80\beta_c$. The two curves 
are fits to the Ansatz (\ref{eq:fit_2}) with
$\xi_d = 2.7232(35)$ ($L \times L$) and $\xi_d=2.7275(24)$ ($2L \times L$), 
in excellent agreement with the exact result 
$\xi_d^{\rm theory} = 2.728865\dots$.
  \end{description}
%
%
%
\clearpage
\newpage
\addtolength{\topmargin}{3cm}
\addtolength{\textheight}{3cm}
%
%
%
\begin{figure*}[htb]
\vskip 5.5truecm
\includegraphics{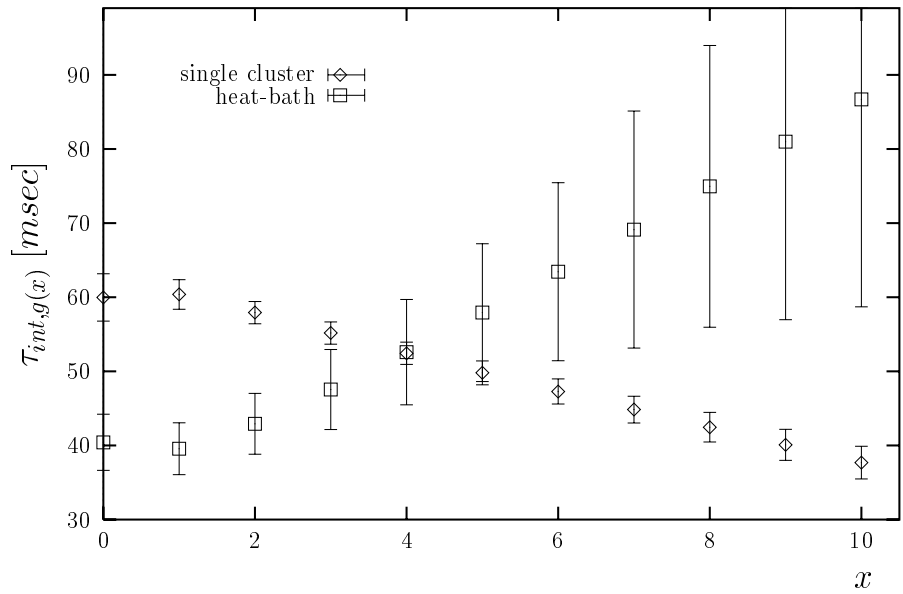}
\caption{\label{fig:auto}Integrated autocorrelation times of $g(x)$ for $q=20$.}
\end{figure*}
\clearpage
\newpage
\addtolength{\topmargin}{-3cm}
\addtolength{\textheight}{-3cm}
%
%
%
\begin{figure*}[htb]
\vskip 16.5truecm
\includegraphics{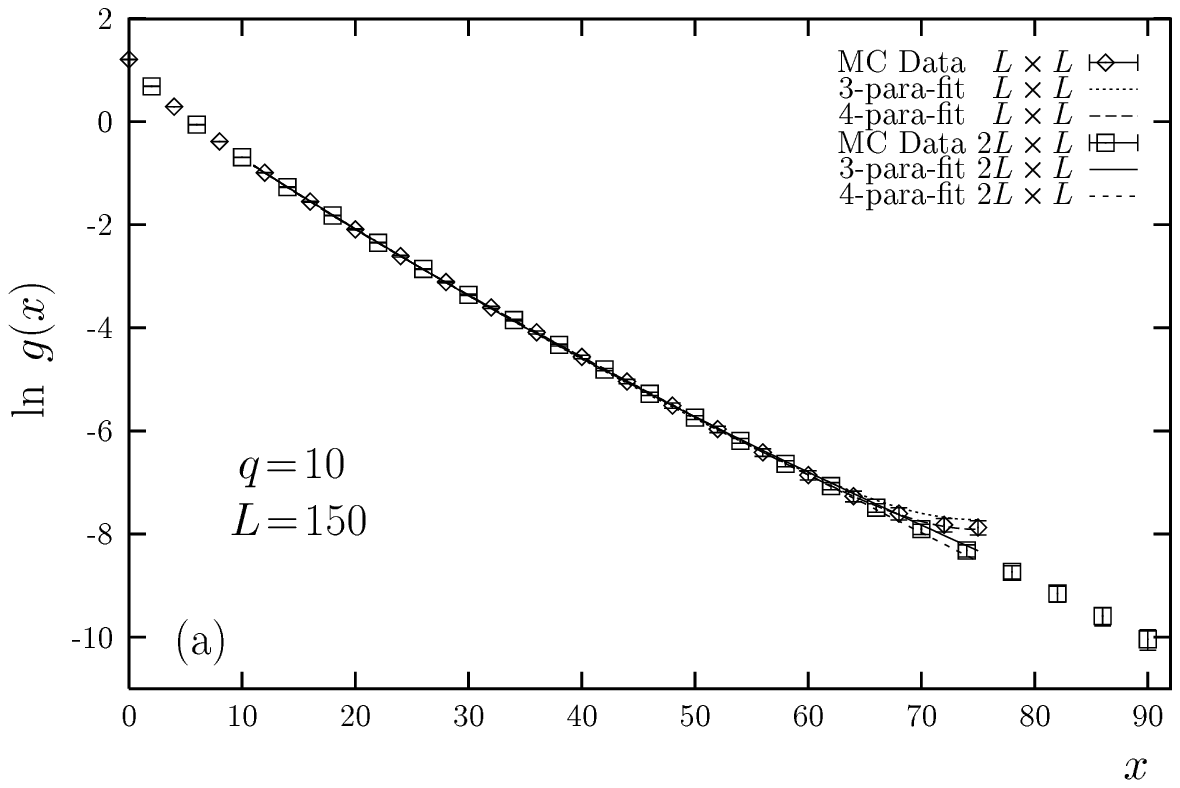}
\includegraphics{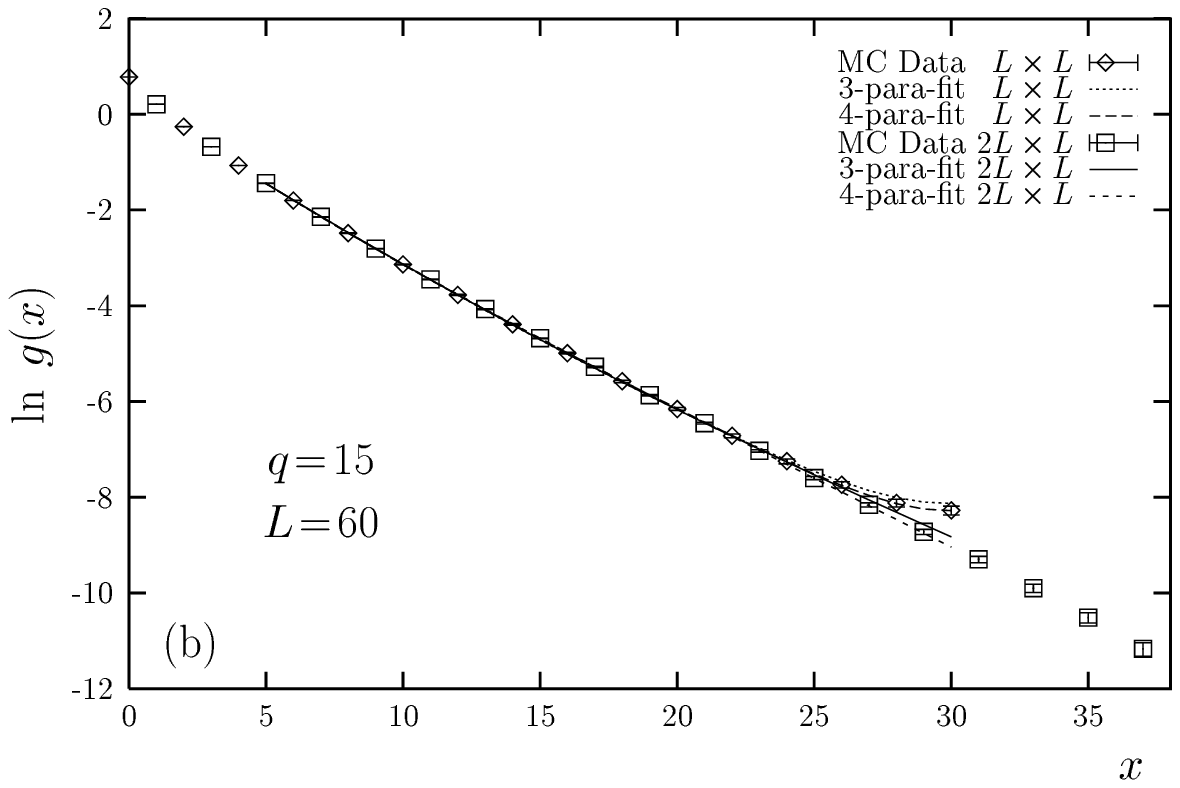}
\includegraphics{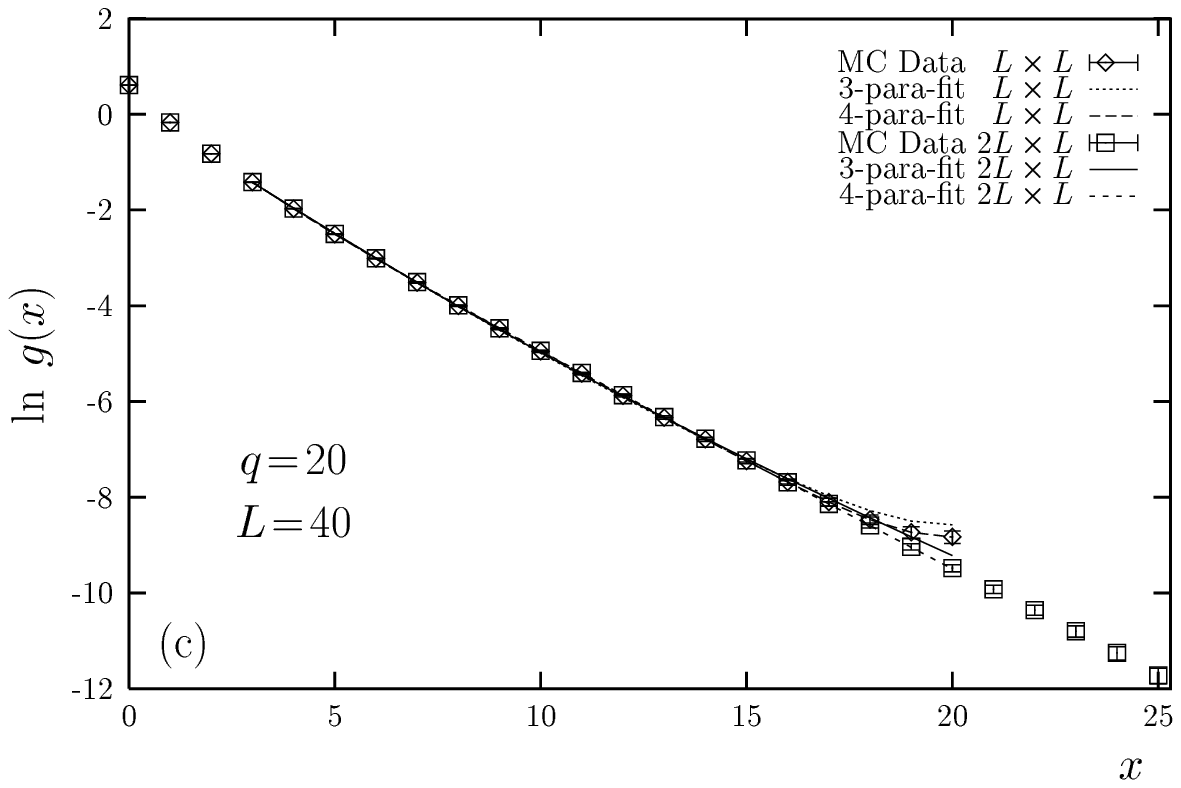}
\caption[a]{\label{fig:fits}
Semi-log plots of the correlation functions $g(x)$ vs distance $x$ on
$L \times L$ and $2L \times L$ lattices for
(a) $q=10$, (b) $q=15$, and (c) $q=20$. The solid and dotted lines are
three-parameter fits to the Ansatz (\ref{eq:fit_4}) with $\xi_d$ held fixed
at its theoretical value. The short and long dashed lines show unconstrained
four-parameter fits over the same $x$ range. For clarity some data points are
discarded.}
\end{figure*}
%
%
\clearpage
\newpage
\addtolength{\topmargin}{3cm}
\addtolength{\textheight}{3cm}
\begin{figure*}[htb]
\vskip 15.5truecm
\includegraphics{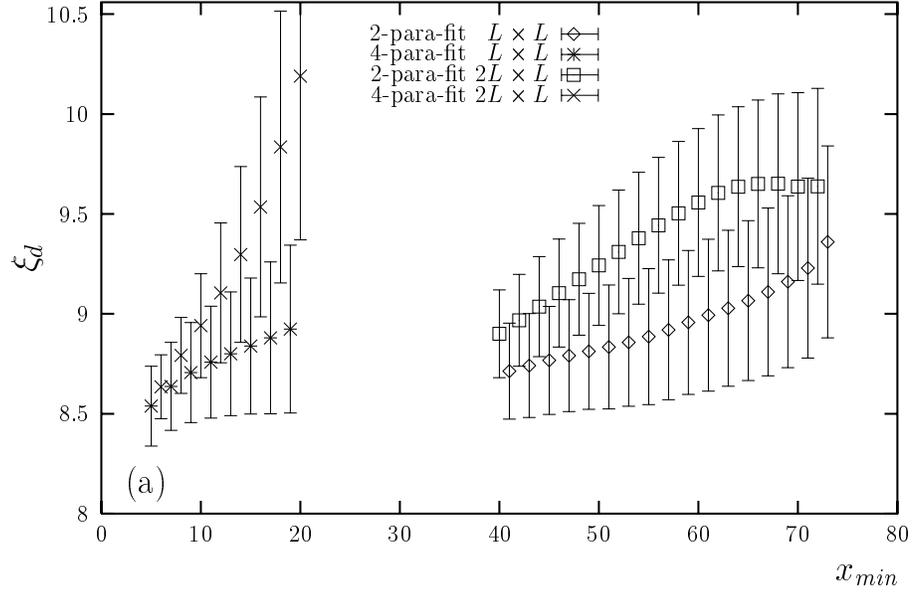}
\includegraphics{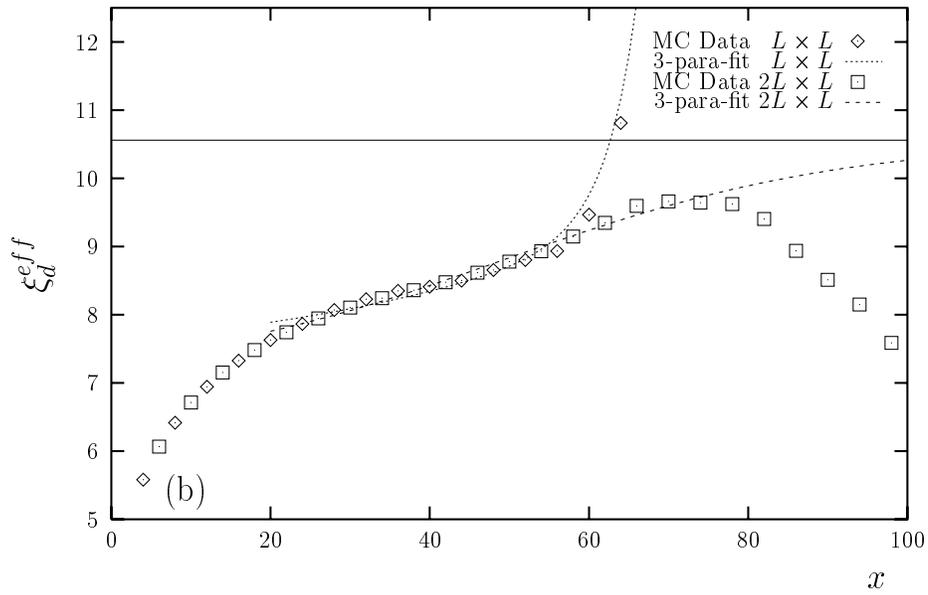}
\caption[a]{\label{fig:xi_eff}
(a) Results for $\xi_d$ of the various fits for $q=10$ using all
data points with $x_{\rm min} \le x \le x_{\rm max}=L/2$ as a function
of $x_{\rm min}$.\\
(b) The effective correlation length (\ref{eq:xi_eff}) vs distance for $q=10$.
The dashed lines are constrained three-parameter fits to the data and the
horizontal line shows the theoretically expected result for $\xi_d$.}
\end{figure*}
%
%
%
\clearpage
\newpage
\addtolength{\topmargin}{-3cm}
\addtolength{\textheight}{-3cm}
\begin{figure*}[htb]
\vskip 8.5truecm
\includegraphics{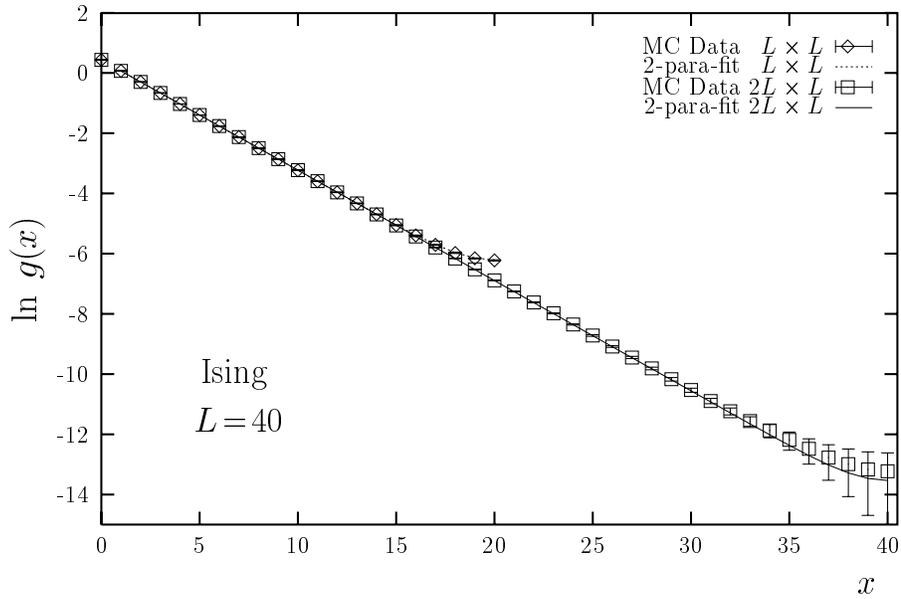}
\caption[a]{\label{fig:isi}
Semi-log plot of the correlation function $g(x)$ of the
2D Ising model at $\beta = 0.71 \approx 0.80\beta_c$. The two curves
are fits to the Ansatz (\ref{eq:fit_2}) with
$\xi_d = 2.7232(35)$ ($L \times L$) and $\xi_d=2.7275(24)$ ($2L \times L$),
in excellent agreement with the exact result
$\xi_d^{\rm theory} = 2.728865\dots$.}
\end{figure*}
\end{document}